\documentclass{PoS}
\usepackage{graphicx}
\usepackage[utf8]{inputenc}
\usepackage{wrapfig}

\title{The International Cosmic Day --\\An Outreach Event for Astroparticle Physics}

\ShortTitle{The International Cosmic Day}

\author{M. Hütten$^{a,b}$, \speaker{T. Karg}$^{,a}$, C. Schwerdt$^{a}$, C. Steppa$^{a}$ and M. Walter$^{a}$\\
        $^{a}$DESY, Platanenallee 6, 15738 Zeuthen, Germany\\
        $^{b}$Humboldt-Universität zu Berlin, Unter den Linden 6, 10099 Berlin, Germany\\
        E-mail: \email{carolin.schwerdt@desy.de}}

\abstract{The International Cosmic Day (ICD) is an astroparticle physics outreach event for high-school students and brings together students and different physics outreach projects from all over the world. Groups of scientists, teachers, and students meet for one day to learn about cosmic rays and perform an experiment with atmospheric muons. All participating groups investigate an identical question. The students are enabled to work together like in an international collaboration, discussing their results in joint video conferences. Analyzing data, comparing and discussing with other ``young scientists'' gives the students a glimpse of how professional scientific research works. Scientists join the video conferences and give lectures to provide an insight in current astroparticle physics research. Several participating research experiments analyze big science data tailored to the questions addressed by the students and present their results on equal terms with the students. To create a lasting event, proceedings with measurement results of all participating groups are published. Every participant receives a personal e-mail with his certificate and the proceedings booklet. Organized by DESY in cooperation with Netzwerk Teilchenwelt, IPPOG, QuarkNet, Fermilab, and national partners like INFN, the ICD is a growing event with more and more popularity. We present the organization of the event and the experience from five years of ICD.}

\FullConference{35th International Cosmic Ray Conference --- ICRC2017\\
		10--20 July, 2017\\
		Bexco, Busan, Korea}

\begin{document}

\section{Introduction}
\label{sec:introduction}

The 2012 centennial of the discovery of cosmic rays motivated us to organize the first International Cosmic Day (ICD). Questions like

\begin{itemize}
\item What are the properties of cosmic rays?
\item What are the sources of cosmic rays?
\item How can cosmic accelerators produce particles with energies several orders of magnitude higher than those reached by a man-made particle accelerators, like the LHC at CERN?
\item How can muons be used to confirm the time dilatation of Einstein's theory of relativity?
\item Is it possible that cosmic radiation influences cloud formation, weather conditions, and even the climate?
\item Have cosmic rays had any influence on the evolution of life on Earth?
\end{itemize}

\noindent are not only interesting to scientists, but also to high school students all over the world. During the ICD, students discuss these questions and find answers to some of them. They carry out measurements and get in contact with other students all around the world to compare and discuss their results. For one day, they work like scientists in an international collaboration.

\section{Target Audience}
\label{sec:audience}

The ICD connects students, teachers, and scientists to learn and discuss about cosmic rays. The students work together in groups supervised by a teacher or a scientist. Teachers with access to cosmic ray detectors can participate with their classes. Scientists engaged in outreach projects can visit schools or invite students to their labs. 

\paragraph{Students} learn about scientific methods: carrying out an experiment, interpreting the data, comparing measurement results of the same physical quantity measured with different detectors and detector types. They are enabled to ask their own questions, make hypotheses and test them in experiments. They get an insight into a current research topic and can make contact to scientific institutions. The students also get in contact with fellow pupils worldwide and have the possibility to practice communicating in English during the common video conference.

\paragraph{Teachers} can use the ICD as a first step in the transfer of knowledge about modern physics, adding topics of particle and astroparticle physics, cosmology and special relativity to their curriculum. These topics can also be studied with interested groups of students outside of the normal courses. The teachers make contact to scientific institutions and other teachers with similar interests.

\paragraph{Scientists} have an opportunity to share their fascination with astroparticle physics with a wide range of students. The ICD enables networking between projects developing and providing cosmic ray experiments for schools. It is an outreach platform for scientists and a platform for experiments to reach a wide audience.

\section{Educational Objectives}
\label{sec:objectives}

At the ICD all participating groups focus on one common question: the zenith angle distribution of atmospheric muons. They try to answer it with an experiment that can be performed and analyzed in one day. The students are given the tasks: 

\begin{itemize}
\item What do you think: Is the number of air shower particles arriving from the horizon the same as from the zenith?
\item Measure the rate of air shower particles for different directions.
\item Interpret your observations.
\end{itemize}

\subsection{Physics}

\begin{wrapfigure}{R}{0.5\textwidth}
\centering
\includegraphics[width=0.48\textwidth]{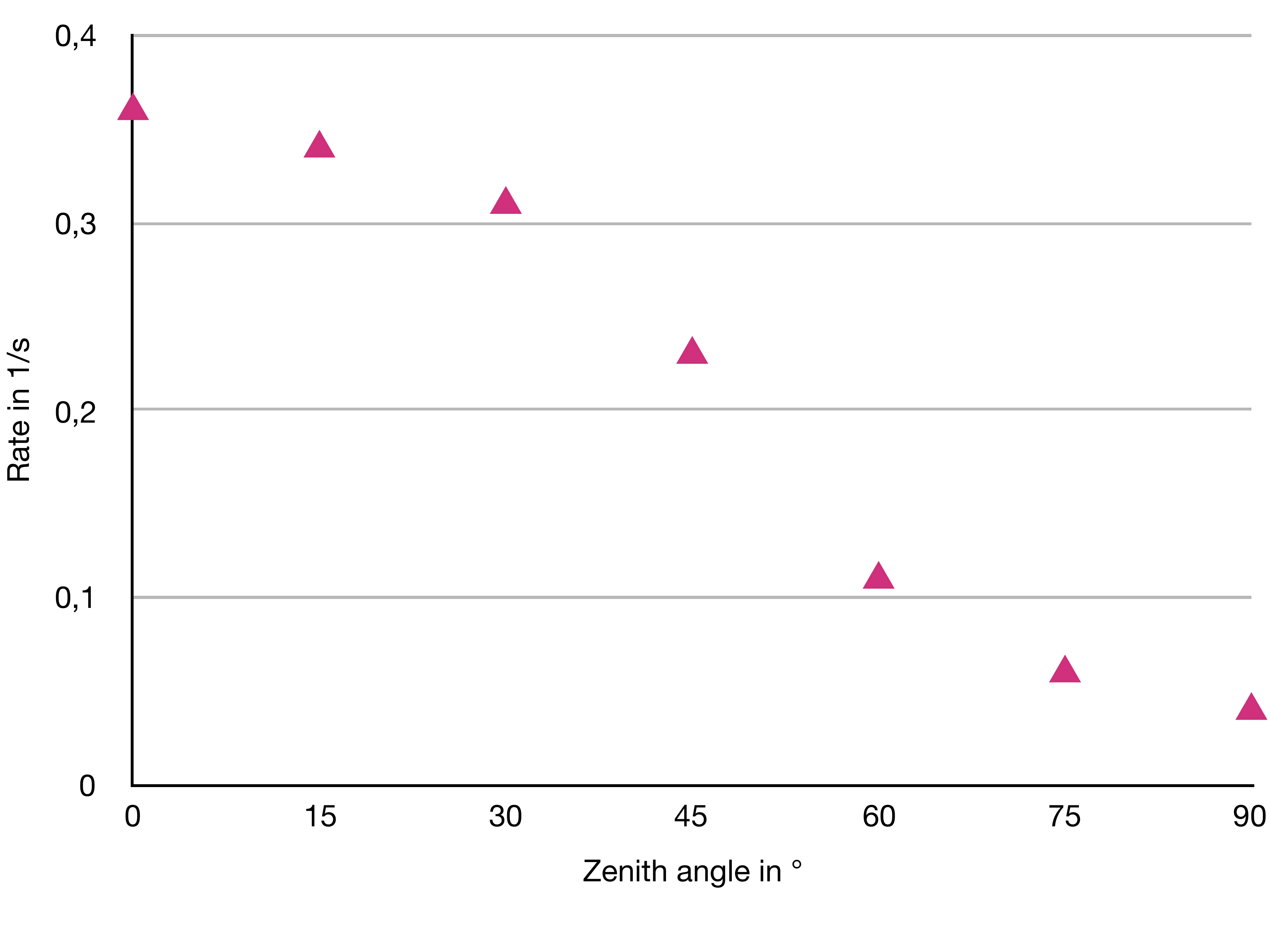}
\caption{Zenith angle distribution of atmospheric muons measured by students with the CosMO detector during the 2016 ICD.}
\label{fig:zenith_angle}
\end{wrapfigure}

The relatively young research field of astroparticle physics has been developing dynamically over the last years, forming a new field between particle physics, astrophysics, and cosmology. For students, it demonstrates that the combination and connection of different fields of science are essential for the investigation of new phenomena as well as for their scientific interpretation.

During the ICD, the students get to know that the universe is filled with different particle accelerators. They learn that one of the crucial topics in astroparticle physics is trying to understand the acceleration mechanisms of cosmic ray particles up to very high energies, much higher than accelerators on Earth can reach. They find out that when high-energy cosmic rays impact on the Earth's upper atmosphere, extensive air showers are initiated. The air showers create scores of secondary particles which can be measured by the students. 

The goal of the ICD is to investigate the zenith angle dependence of atmospheric muons (see Fig.~\ref{fig:zenith_angle}). With two scintillation counters in coincidence it can be measured by all participating groups. The main question that has to be discussed and answered is why the muon rate has its minimum if the detectors are oriented horizontally. This leads to the lifetime of muons and their decay. Muon production and their propagation through the atmosphere can be discussed.

\subsection{Experiments for Students}

\begin{wrapfigure}{R}{0.5\textwidth}
\centering
\includegraphics[width=0.48\textwidth]{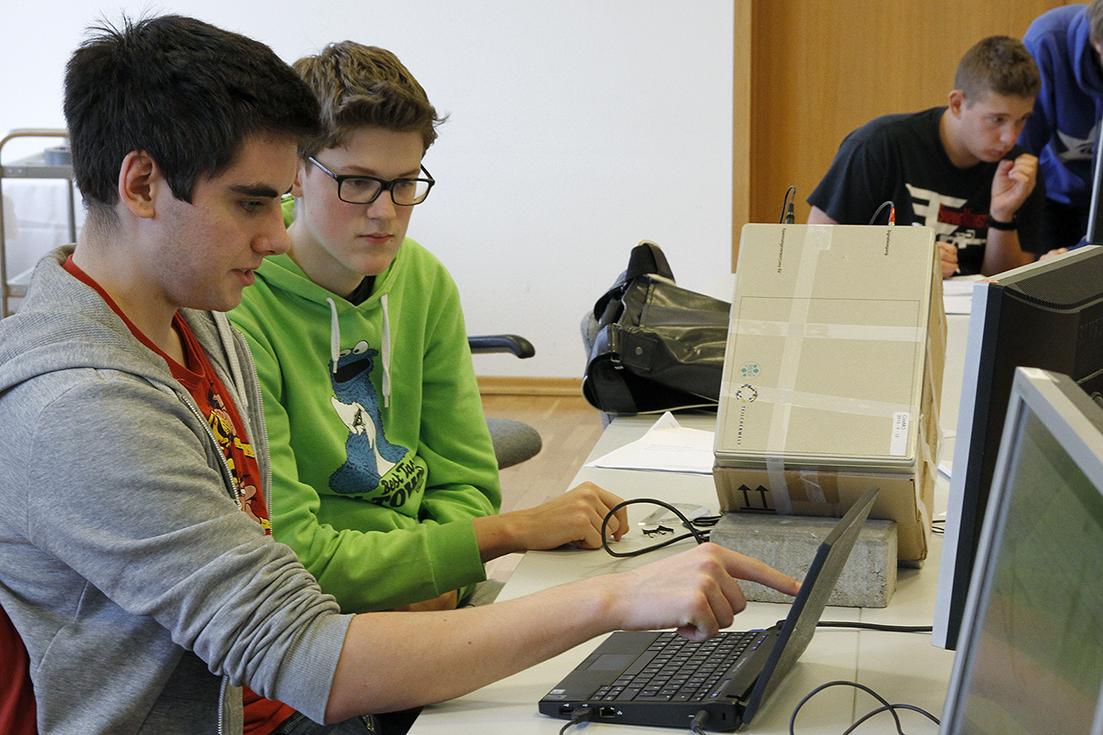}
\caption{Students at the ICD in 2015 performing a measurement of the zenith distribution of atmospheric muons using the CosMO detector \cite{cosmo}.}
\label{fig:students}
\end{wrapfigure}

The detectors used for the ICD apply similar techniques like large-scale scientific experiments. Most student experiments are based on scintillation counters. These are employed e.g.~in the German \textit{CosMO-Experiment} \cite{cosmo}, in \textit{QuarkNet} in the U.S.~\cite{QuarkNet} and in \textit{cosmodétecteur} developed by Sciences à l'École in France \cite{France}. Using a data acquisition (DAQ) board and software specifically developed for students, the participants can set up and calibrate the detectors, and perform measurements on their own. Since the DAQ board supports coincidence triggers between multiple scintillators, the detectors can be operated as muon hodoscopes enabling direction-dependent muon flux measurements (Fig.~\ref{fig:students}). 

The CosMO counters consists of plastic scintillators with a size of $20 \times 20 \times 1.2 \, \textrm{cm}^3$. At least two counters are used in coincidence to suppress electronic noise and are placed in parallel to each other at a mutual distance $d$, which defines the solid angle from which particles are accepted. It has been shown that distances between 20 and 30\,cm give a good compromise between restriction to direction of arrival and too low rates. 

Principally, participating at the ICD is possible with any measuring device sensitive to atmospheric muons. Besides scintillation detectors, so-called \textit{Kamiokanne} counters \cite{kamiokanne} have been used by several groups in the past. The Kamiokanne, a blend word of ``Kamiokande'' and ``Kanne'', the German word for pot, is a standard one liter thermos flask with an inner reflective coating as part of the insulation. Filled with common tap water, traversing atmospheric muons cause a sufficiently bright Cherenkov light flash to be detected by a photomultiplier tube (PMT) mounted in the nozzle instead of the lid. The PMT signal can be read out with the same DAQ board and software used with scintillation counter detectors. Like the scintillation counters, several Kamiokanne detectors can be linked to establish a trigger condition, however, due to the pots' geometry and the lower event rates, the determination of the directionally varying muon flux is much more imprecise than when using flat scintillation panels.

As for all experiments, an essential element of scientific work is the analysis and interpretation of the data. New for most of the students is the amount of data and their handling using a computer. These big data sets offer many possibilities to develop and investigate new ideas and to apply the students school knowledge in a new context. 

\subsection{Large-Scale Experiments}

A novelty in 2016 was that PhD students from the large-scale experiments ATLAS \cite{ATLAS} and IceCube \cite{icecube} presented their atmospheric muon data during a video conference at the ICD. The participating student groups got an impression of how these large experiments are built up and what the most important detector components are in comparison to their own small-scale experiments. ATLAS is one of the four big detectors at the Large Hadron Collider (LHC) at CERN, located in a cavern 100 m below ground. IceCube is the world’s largest neutrino detector and is located at the South Pole. Encompassing a cubic kilometer of ice, it comprises of 86 strings, each instrumented with 60 photosensors. In both experiments atmospheric muons are a background signal detected with high rate although most of the muons are filtered by the thick layer of rock or ice. During the video conference the muon rates were shown in dependence of the zenith angle also for the ATLAS and IceCube experiments. The comparison with the students measurements resulted in a fruitful discussion of the interpretation of the different results.

\section{Implementation of the International Cosmic Day}
\label{sec:implementation}

\subsection{Web Page}

\begin{figure}[t!]
\centering
\includegraphics[width=\textwidth]{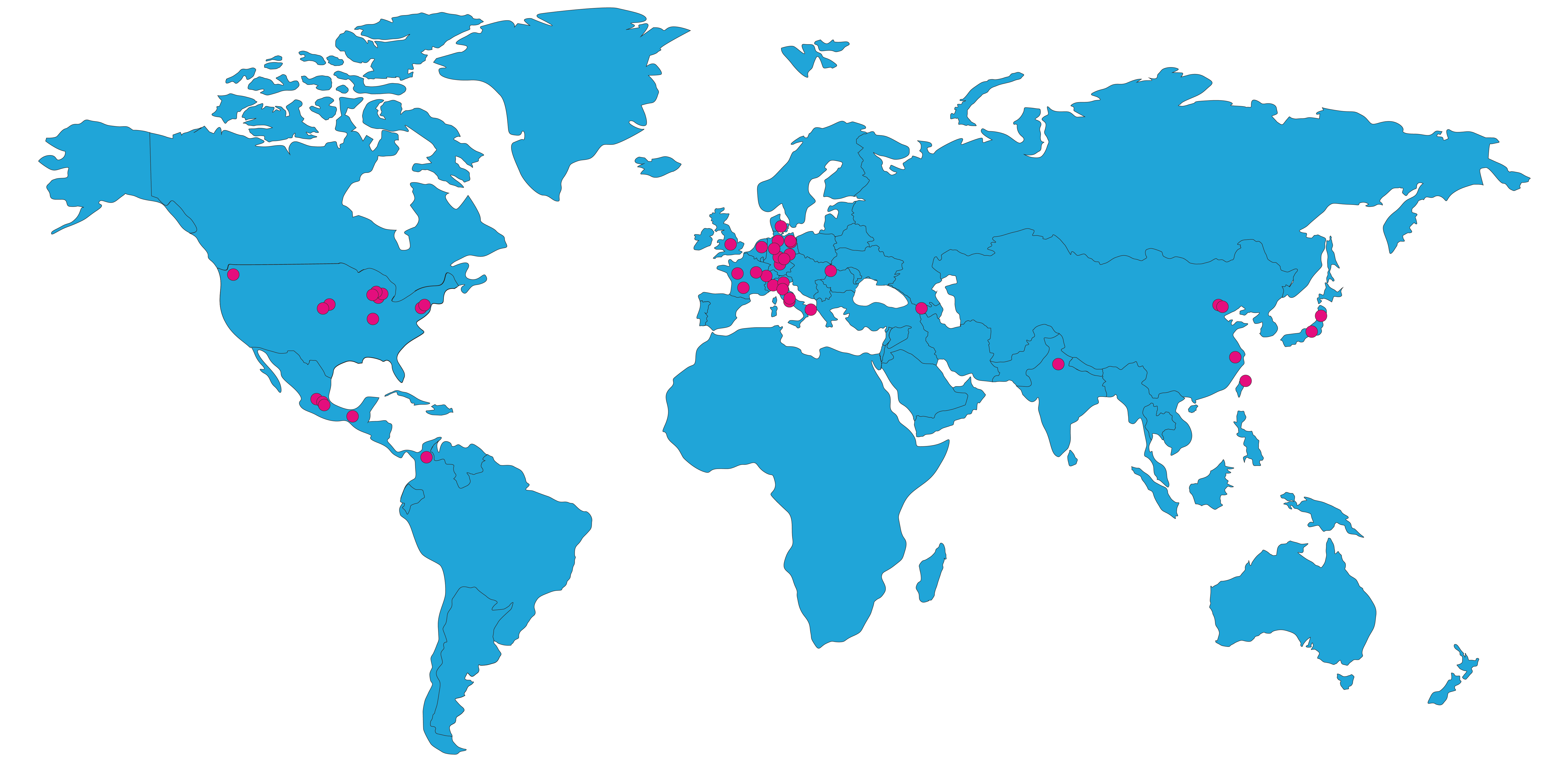}
\caption{Locations of all groups participating at the ICD in 2016.}
\label{fig:worldmap}
\end{figure}

The web page \href{https://icd.desy.de}{https://icd.desy.de} is the central information platform for the ICD. It contains general information on the ICD like its goals and a compilation of experiments for students. There is also special information for each year's Cosmic Day.
The page was designed with a focus on the three target groups students, teachers, and scientists. Further, the page contains a map displaying the locations of all participating groups (Fig.~\ref{fig:worldmap}). In addition, there is a Facebook page\footnote{\href{https://www.facebook.com/InternationalCosmicDay/}{https://www.facebook.com/InternationalCosmicDay/}} for connecting the participating groups and for sharing information quickly and easily.

\subsection{Advertising}
We have compiled a database of scientists interested in astroparticle physics-related student projects and of teachers who have completed some additional training in astroparticle physics. We contact them about six months before the ICD and provide them with announcement posters and press releases that can be localized. The ICD is also advertised via different science platforms, e.g.~APPEC, Welt der Physik, and IPPOG.

\subsection{Participation}
One of the main purposes of the ICD is to encourage collaborative work. The participants are supposed to work in groups supervised by an experienced teacher or scientist and can only register as such. The supervisor is responsible for inviting and keeping contact to the students and for the local organization. The event can be held at a school, at a university or lab, or also at other venues like e.g.~museums. A further requirement is access to some sort of cosmic ray detector.

\subsection{Schedule}
Each participating group plans their own day. In general, the following agenda items will be part of the day: an introduction to cosmic rays, the measurement of cosmic particles, in particular the zenith angle distribution of atmospheric muons, data analysis, discussion of the results within the group and with other groups worldwide via video conferences, and presentation of the results in the ICD booklet.

\subsubsection{Introductory Presentation}
The introductory presentation should be given by a scientist who will choose the appropriate level for the talk. It can cover topics such as:

\begin{itemize}
\item What are cosmic rays?
\item Who discovered them and how?
\item What are the open questions in astroparticle physics today and how are they addressed?
\item What experiment will the local group carry out and how does it relate to the questions above?
\end{itemize}

The presentation can either be given at the beginning of the day or it can be split in one part in the morning and one part in the afternoon. It can also be complemented by more popular content, e.g.~the topic ``IceCube -- Wintering at the South Pole''.

\subsubsection{Measurement and Data Analysis}
Each group will individually plan and carry out the measurement. An extension of this measurement in order to investigate other aspects of extensive air showers is also possible, e.g. the measurement of the velocity and mean lifetime of muons. There are no restrictions on the schedule or on the setup used. For the international exchange it is important that at the end of the day the results are summarized and published in a diagram like the one shown in Fig.~\ref{fig:zenith_angle}. The groups are encouraged to provide additional information on their detector setup and measurements.

\subsubsection{Discussion of Measurement Results}
Since one of the goals of this project is to allow school students to compare and discuss their results with other groups worldwide, we have created a timetable listing participating institutes in nearby timezones. 

A text chat is provided for students to get in contact with other groups and discuss problems or preliminary results. We have observed that the chat is used by about half of the participating groups, mainly to ask questions and share photos of their project.

Furthermore, several video calls are organized for the different time zones. The calls are held in English. The goal of the conferences is that each group presents their detector and measurement results to their peers. To motivate the purpose of the calls, the organizers are encouraged to explain to their students how scientific collaborations work together in practice and how they exchange their results.

\subsubsection{Closeout and Sustainability}
To create a long-lasting effect, all participating groups are tasked to document their results with photos, notes, and their measurement result on one page. The organizing school or institution can present itself on an additional page. All contributions, together with some introductory material, are collected in a booklet, similar to conference proceedings. Each participant receives a copy of the booklet together with a certificate. The design of the booklet allows the students iterating the topic of the ICD at home and compare their results to other group's results. Links and references to school projects are provided. The teachers can utilize the booklet for preparation of lessons. 

\section{Facts and Figures}

\begin{figure}[t!]
\centering
\includegraphics[width=0.9\textwidth]{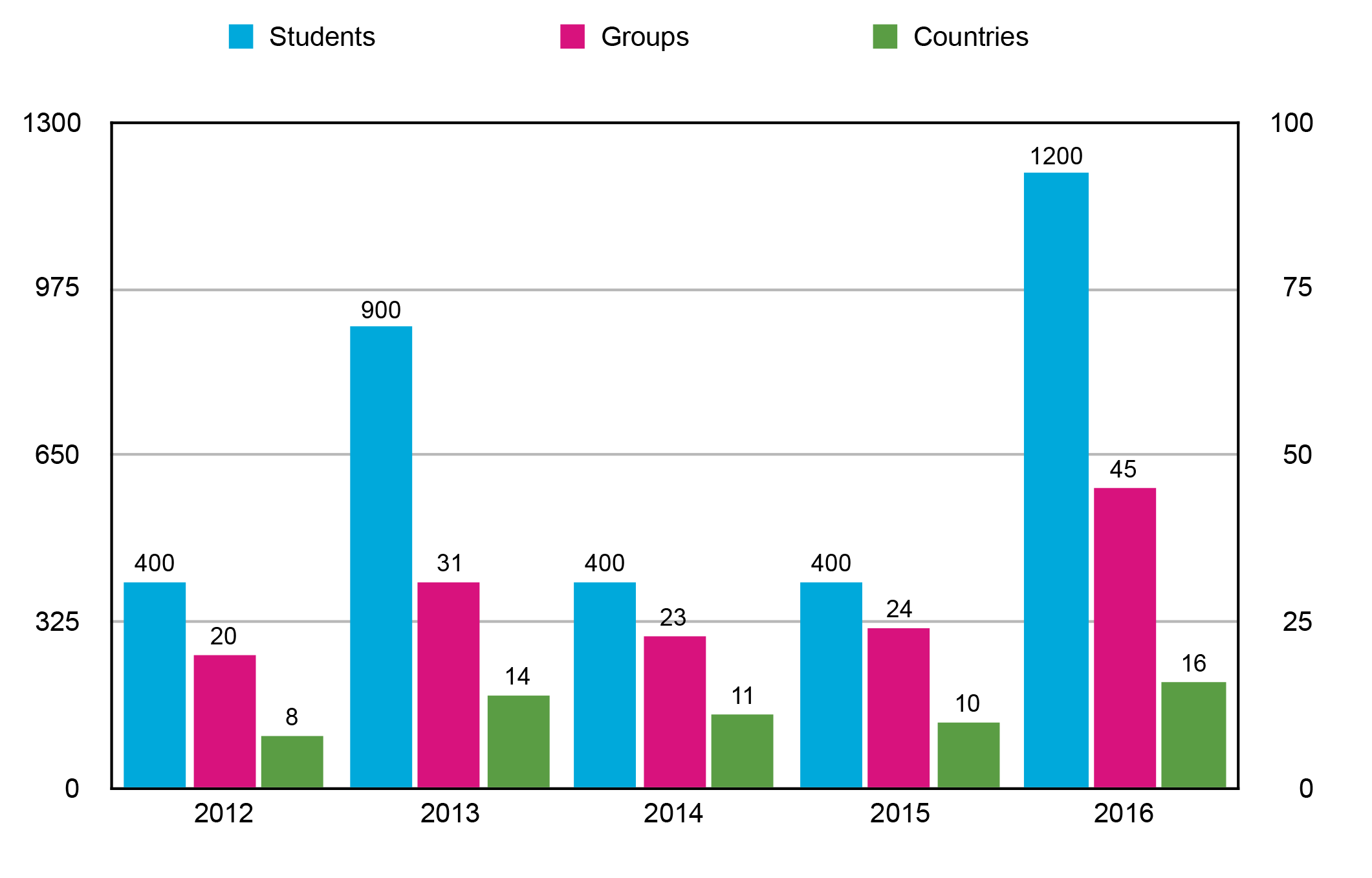}
\caption{Number of students, groups, and countries participating in the ICD since its establishment in 2012.}
\label{fig:participation}
\end{figure}

Since 2012, about 3300 students from 26 countries have participated at the five previous ICDs. In Fig.~\ref{fig:participation}, the participation is shown separately for each year. Overall, it can be seen that the number of students per participating group remained approximately constant over the last years, with about 20 to 30 students per group. We experienced that the event works best in smaller groups with 20 or less students. Typically, every group has access to only one or two detector sets, since the costs per set is relatively high for typical public school budgets. Ideally, two students share one detector set and experiment autonomously. Larger group sizes per detector are still feasible, however, experience showed that not more than six students should share a detector for autonomous experimenting, and experiments should be demonstrated by the teacher in larger groups.

Figure~\ref{fig:worldmap} shows that in last year, most countries with well established student outreach programs for cosmic-ray experiments have participated in the ICD. There are many more countries involved in large-scale cosmic ray research but which lack access to detectors that would enable students to participate in the ICD.

\section{Outlook}

To enable more students and teachers to participate at the ICD, we plan to provide measurement data in a suitable format to be analyzed by students without access to experiments at the ICD. The data set should be large enough to leave room for students to realize their own analysis ideas and make discoveries. However, the larger the data set becomes, the more documentation needs to be read by the student before starting an analysis and the more difficult it is to work with the data using the tools available in a high school environment. The threshold for the students to start an analysis increases quickly to the point where working with this data becomes unattractive. To prevent this, tools need to be created that allow students accessing large and unfiltered data in an intuitive way without requiring long special training. First experience has been gained e.g.~within the International Masterclass project \cite{IntMC} for accelerator-based particle physics and within Cosmic@Web \cite{Cosmic@Web} for cosmic ray data, which currently is only available in German. Following these examples we plan to develop a similar web-based tool for the ICD.

In summary, the International Cosmic Day has been established as an event that is valued by many local organizers who join it with groups of students every year. We plan to continue organizing it on a yearly basis; the next ICD is scheduled for November 2017.


\begin{thebibliography}{99}
\bibitem{cosmo}
R. Franke et al.,
\emph{CosMO -- A Cosmic Muon Observer Experiment for Students},
in Proceedings of the \emph{33rd International Cosmic Ray Conference},
{\tt arXiv:1309.3391 [astro-ph.IM]}.
\bibitem{QuarkNet}
\href{https://quarknet.i2u2.org}{https://quarknet.i2u2.org}
\bibitem{France}
\href{http://www.sciencesalecole.org/plan-cosmos-a-lecole-materiel/}{http://www.sciencesalecole.org/plan-cosmos-a-lecole-materiel/}
\bibitem{kamiokanne}
\href{http://kamiokanne.uni-goettingen.de}{http://kamiokanne.uni-goettingen.de}
\bibitem{ATLAS}
ATLAS Collaboration,
%% \emph{The ATLAS experiment at the CERN Large Hadron Collider},
\emph{JINST} {\bf 3} (2008) S08003.
\bibitem{icecube}
M.~G.~Aartsen et al. (IceCube Collaboration),
%% \emph{The IceCube Neutrino Observatory: instrumentation and online systems},
\emph{JINST} {\bf 12} (2017) P03012.
\bibitem{IntMC}
\href{http://www.physicsmasterclasses.org}{http://www.physicsmasterclasses.org}
\bibitem{Cosmic@Web}
\href{http://cosmicatweb.desy.de/ctplot/}{http://cosmicatweb.desy.de/ctplot/}
\end{thebibliography}
\end{document}